\documentclass[fleqn,usenatbib]{mnras}

\usepackage{newtxtext,newtxmath}
\usepackage[T1]{fontenc}
\usepackage{ae,aecompl}
\hypersetup{draft}

\usepackage{graphicx}	% Including figure files
\usepackage{amsmath}	% Advanced maths commands
\usepackage{amssymb}	% Extra maths symbols

\usepackage{BibDef}
\usepackage{bm}
\usepackage{ulem}

\title[CRs in dense cores]{A new proxy to estimate the cosmic-ray ionisation rate in dense cores}

\author[S. Bovino et al.]{
S. Bovino$^{1}$\thanks{E-mail: stefanobovino@udec.cl},
S. Ferrada-Chamorro$^1$, A. Lupi$^2$, D.~R.~G.~Schleicher$^{1}$, and P. Caselli$^{3}$\\
$^{1}$Departamento de Astronom\'ia, Facultad Ciencias F\'isicas y Matem\'aticas, Universidad de Concepci\'on, \\
Av. Esteban Iturra s/n Barrio Universitario, Casilla 160, Concepci\'on, Chile\\
$^{2}$Scuola Normale Superiore, Piazza dei Cavalieri 7, Pisa, IT-56126 Italy\\
$^{3}$Centre for Astrochemical Studies, Max-Planck-Institute for Extraterrestrial Physics, Giessenbachstrasse 1, 85748, Garching, Germany
}

\date{Accepted XXX. Received YYY; in original form ZZZ}

\pubyear{2019}

\begin{document}
\label{firstpage}
\pagerange{\pageref{firstpage}--\pageref{lastpage}}
\maketitle

% Abstract of the paper
\begin{abstract}
Cosmic rays are a global source of ionisation, and the ionisation fraction represents a fundamental parameter in the interstellar medium. Ions couple to magnetic fields, affect the chemistry, and the dynamics of star-forming regions as well as planetary atmospheres. However, the cosmic-ray ionisation rate represents one of the bottlenecks for astrochemical models, and its determination is one of the most puzzling problems in astrophysics. While for diffuse clouds reasonable values have been provided from H$_3^+$ observations, for dense clouds, due to the lack of rotational transitions, this is not possible, and estimates are strongly biased by the employed model. We present here an analytical expression, obtained from first principles, to estimate the cosmic-ray ionisation rate from observational quantities. The theoretical predictions are validated with high-resolution three-dimensional numerical simulations and applied to the well known core L1544; we obtained an estimate of $\zeta_2 \sim 2-3 \times 10^{-17}$ s$^{-1}$. Our results and the analytical formulae provided represent the first model-independent, robust tool to probe the cosmic-ray ionisation rate in the densest part of star-forming regions (on spatial scales of $R \leq 0.05$ pc). An error analysis is presented to give statistical relevance to our study.
\end{abstract}

% Select between one and six entries from the list of approved keywords.
% Don't make up new ones.

%%%%%%%%%%%%%%%%%%%%%%%%%%%%%%%%%%%%%%%%%%%%%%%%%%
\def\lsim{\mathrel{\rlap{\lower 3pt \hbox{$\sim$}} \raise 2.0pt \hbox{$<$}}}
\def\gsim{\mathrel{\rlap{\lower 3pt \hbox{$\sim$}} \raise 2.0pt \hbox{$>$}}}
\def\msun{\rm {M_\odot}}
\def\kms{\rm km\,s^{-1}}
\def\mach{\mathcal{M}}
\def\angstrom{\mathring{\mathrm{A}}}
\def\grav{\rm G}
\def\gizmo{\textsc{gizmo}}

\def\al#1{\textcolor{red}{[AL: #1]}}
%%%%%%%%%%%%%%%%% BODY OF PAPER %%%%%%%%%%%%%%%%%%

\section{Introduction}
%Star formation is strongly affected by the interplay between microphysical processes and dynamics. In this exchange of energy, coupling motions, and dynamical changes, chemistry plays a crucial role.  Typical conditions of molecular clouds drive the formation of specific molecules and the consequent chemical processes. 
The densest regions of molecular clouds where star-formation is beginning are characterized by a high degree of CO depletion \citep[e.g.][]{Caselli1999,Fontani2012,Giannetti2014,Sabatini2019} and high levels of deuteration \citep[e.g.][]{Ceccarelli2014}. In particular, the deuterated forms of the main ion H$_3^+$ (e.g. H$_2$D$^+$) dominate the chemistry during the early stages prior to the formation of a protostellar object \citep{Pagani1992,Caselli2003,Giannetti2019}
%\citep{Pagani1992,Caselli2003,Walmsley2004,Giannetti2019}. 

The formation of H$_3^+$ and more generally the ion-neutral chemistry characterizing these regions, are driven by cosmic rays (CRs), i.e. highly energetic particles that can penetrate the dense regions within molecular clouds starting a chain of reactions which leads to the formation of key tracers (e.g. HCO$^+$, N$_2$H$^+$). H$_3^+$ is indeed considered a key molecule to determine the cosmic-ray ionisation rate (CRIR), due in particular to the simple chain of reactions involved in its chemistry. Each ionisation of H$_2$ through CRs leads to the formation of  H$_3^+$, which is mainly destroyed by neutrals \citep[e.g. CO and O,][]{Dalgarno2006}.

%forms from the followed by reaction with molecular hydrogen

%\begin{eqnarray}\label{eq:reac}
%	\mathrm{H_2} + \zeta_2 & \rightarrow & \mathrm{H_2^+} + e\\
%	\mathrm{H_2^+}  + \mathrm{H_2} & \rightarrow & \mathrm{H_3^+} + \mathrm{H}
%\end{eqnarray}

CRs have a strong effect on the deuteration process itself, this being affected by H$_3^+$, the ion which starts the deuteration process by forming H$_2$D$^+$. Many authors have shown, through simple \citep[e.g.][]{Caselli2008,Kong2015} or complex theoretical models \citep{Koertgen2018,Bovino2019}, that a higher CRIR favours the deuteration process, in particular shortening the time to reach typical observed values. Determining the timescale for deuteration is of fundamental importance when the deuterium fractionation is used as a chemical clock of star-forming regions \citep[see e.g.][]{Fontani2011,2014Nature}, and providing an estimate for the CRIR has then a crucial effect on our interpretation of the physics of star-formation (e.g. ambipolar diffusion, timescales, chemistry).

The CRIR has been probed through a mix of observations and chemical models in diffuse clouds starting from the pioneering work of \citet{Black1978} based on OH and HD observations. Quantitative measurements were then provided through observations of H$_3^+$ in absorption in diffuse and dense clouds by \citet[e.g.][]{Indriolo2012,Neufeld2017} and \citet{Gebelle1996} and \citet{McCall1999}. Additional work was pursued by \citet{vanderTak2000}, with a different approach based on H$^{13}$CO$^+$ in dense clouds towards massive young stars. 
Overall, in particular for dense clouds, estimates of the ionization rate of hydrogen atom $\zeta_\mathrm{H}$ are in the range $\sim 3\times10^{-17}-10^{-16}$ s$^{-1}$. From now on we will consider $\zeta_2 = 2.3 \zeta_\mathrm{H}$ as the CRIR of hydrogen molecule. 

Analytical approaches, based on simple steady-state assumptions, have been explored in different works by \citet{Caselli1998,Caselli2002,Ceccarelli2004,Vaupre2014}. 
These works were focused on the estimate of the ionisation fraction first and then CRIR from DCO$^+$, HCO$^+$, and CO. However, \citet{Shing2016} have shown that estimates based on the ratio  [HCO$^+$]/[DCO$^+$] strongly depend on the initial H$_2$ ortho-to-para ratio and other dynamical quantities like the age of the source, the temperature, and the density.

A pure theoretical method based on dust temperature evaluation has been proposed and applied to L1544 by \citet{Ivlev2019}. The latter provided an estimate of the CRIR of $\zeta_2 \sim10^{-16}$ s$^{-1}$. 
Recently, a new method has been proposed by \citet{Bialy2019}, who suggested to observe line emission of H$_2$ rovibrational transitions to estimate the CRIR. Overall, there is no consense on the final value of $\zeta_2$ in dense regions and estimates vary by orders of magnitude.

%To conclude this historical summary on how the CRIR has been evaluated from different methods \al{, we must \sout{we have also to}} mention estimates based on the ratio  [HCO$^+$]/[DCO$^+$] by \citet{Shing2016}\al{,} who stressed the strong dependence of this method on the initial H$_2$ ortho-to-para ratio and other dynamical quantities like the age of the source, the temperature, and the density.

In this Letter, we aim at providing a robust tool to estimate the CRIR in dense cores from first principles. Our approach is validated by high-resolution numerical simulations and tested under different dynamical and chemical conditions. In the following Sections we shall briefly present the simulations, the methodology, and the error estimates. We then discuss some of the caveats and the applicability of the presented method.

\section{H$_3^+$ from its deuterated forms and CRIR}
Under the steady-state assumption, \citet{Caselli2002} provided useful formulae that can be employed to estimate for instance the ionisation fraction of dense regions. Starting from their Eq. (9) and (10) and the work by \citet{McCall1999} we have built a set of new correlations which can be employed to estimate the CRIR. 
%The formulae are only valid in the presence of substantial amounts of H$_2$D$^+$. 
The entire analysis is based on observable quantities, i.e. column densities. From the observations of H$_2$D$^+$ and other H$_3^+$ isotopologues (enhanced in the cold and dense regions of molecular clouds) we can obtain an estimate of the H$_3^+$ column density:

\begin{equation}\label{eq:h3p}
	N[\mathrm{H_3^+}] = \frac{1}{3}\frac{\mathrm{D}[\mathrm{H_3^+}] }{R_\mathrm{D}}\,,
\end{equation}

\noindent where $R_\mathrm{D}$ is the deuterium fractionation of HCO$^+$, 

\begin{equation}\label{eq:rd}
	R_\mathrm{D} = \frac{N[\mathrm{DCO^+}]}{N[\mathrm{HCO^+}]}\,,
\end{equation}

\noindent and D$[\mathrm{H_3^+}]=N[\mathrm{H_2D^+}] +  N[\mathrm{D_2H^+}] \gamma_1 +  N[\mathrm{D_3^+}] \gamma_2$, with \mbox{$\gamma_1=1-R_\mathrm{D}$} and $\gamma_2 = 2 - R_\mathrm{D}$, and each term represents the sum over the different isomers (ortho, para, and meta). We include all H$_3^+$ isotopologues, to account for the conversion of H$_2$D$^+$ into D$_2$H$^+$, and finally into D$_3^+$. These represent the additional terms compared to the original \citet{Caselli2002} derivation, where we have considered that HCO$^+$ can form also from D$_2$H$^+$, and DCO$^+$ from D$_2$H$^+$ as well as  D$_3^+$. Note that the correction factors are negligible if $R_\mathrm{D} \leq 0.1$.
%\footnote{Typical observed values for L1544 are around 0.01 \citep[][]{Redaelli2019}.}. 
%While we can observe o-H$_2$D$^+$ and p-D$_2$H$^+$ with APEX and ALMA, o-D$_2$H$^+$ and p-H$_2$D$^+$ requires instruments like SOFIA. On the contrary, D$_3^+$ cannot be observed because of a lack of dipole moment. Ideally, 
%This formulation will work better for very early stage sources where we can use just the two tracer observable with ease . In the case of evolved, but still prestellar sources, we have to consider an error or make use of SOFIA to measure the missing isotopologues. 
We test different formulae which gradually include more isotopologues to provide different levels of approximations and an error estimate for the different formulae. We will refer to D$[\mathrm{H}_3^+$] equal to (1) only o-H$_2$D$^+$, (2) the total H$_2$D$^+$ (ortho + para), (3) total H$_2$D$^+$ and total D$_2$H$^+$\footnote{To employ cases (2) and (3) SOFIA observations would be needed as the p-H$_2$D$^+$ and o-D$_2$H$^+$ transitions fall in the THz regime.}, (4) all the isotopologues (i.e. with D$_3^+$)\footnote{Note that D$_3^+$ cannot be observed in high extinction regions as it has transitions in the near infrared, so this formula cannot be applied from an observational point of view but we report it for completeness.}, and (5) only the species observable with ALMA/APEX, i.e. D[H$_3^+$] = o-H$_2$D$^+$ + p-D$_2$H$^+$.

%on the number of tracers  included in the term $D[\mathrm{H_3^+}]$x.

%\paragraph*{The cosmic-ray ionisation rate}

\noindent Once obtained the column density of H$_3^+$, we can have a rough estimate of $\zeta_2$ \citep[see e.g.][]{Oka2019} by balancing formation and destruction of H$_3^+$ (in steady-state):

\begin{equation}\label{eq:crs}
	\zeta_2 = \bar{\alpha} k_\mathrm{CO}^\mathrm{H_3^+} \frac{N[\mathrm{CO}] N[\mathrm{H_3^+}]}{N[\mathrm{H_2}]}\frac{1}{L}
\end{equation}

\noindent with $k_\mathrm{CO}^\mathrm{H_3^+}$ the destruction rate of H$_3^+$ by CO, that we consider the main destruction path for H$_3^+$, $L$ the path length over which the column densities are estimated, and $\bar{\alpha}$ a correction factor which encapsulates any missing effect in the approximation. The factor is calibrated for each of the proposed approximations in the next Sections. We are assuming here that the ratio between number densities is the same as the ratio between column densities, then $X_\mathrm{CO} = n_\mathrm{CO}/n_\mathrm{H_2}= N[\mathrm{CO}]/N[\mathrm{H_2}]$.
Other chemical reactions affecting H$_3^+$ destruction, like for instance destruction via atomic oxygen (assumed to be highly depleted onto dust grains in dense regions), or dissociative recombination with electrons are neglected. The latter represents our strongest assumption but allows us to remove a dependence on the ionisation fraction. However, we have also to consider that $x_\mathrm{e} \leq 10^{-8}$ in the very central region of the cores and then our choice is not going to have a huge impact on the final results. 
%This requires an accurate measurement of CO, which is normally affected by depletion in these regions and can then introduce an additional error.

%In addition, a dependence on the value of $R_D$ needs to be taken into account. When $R_D > 10^{-2}$ we need to apply a correction factor (needs to be investigated).

\begin{figure*}
\includegraphics[scale=0.45]{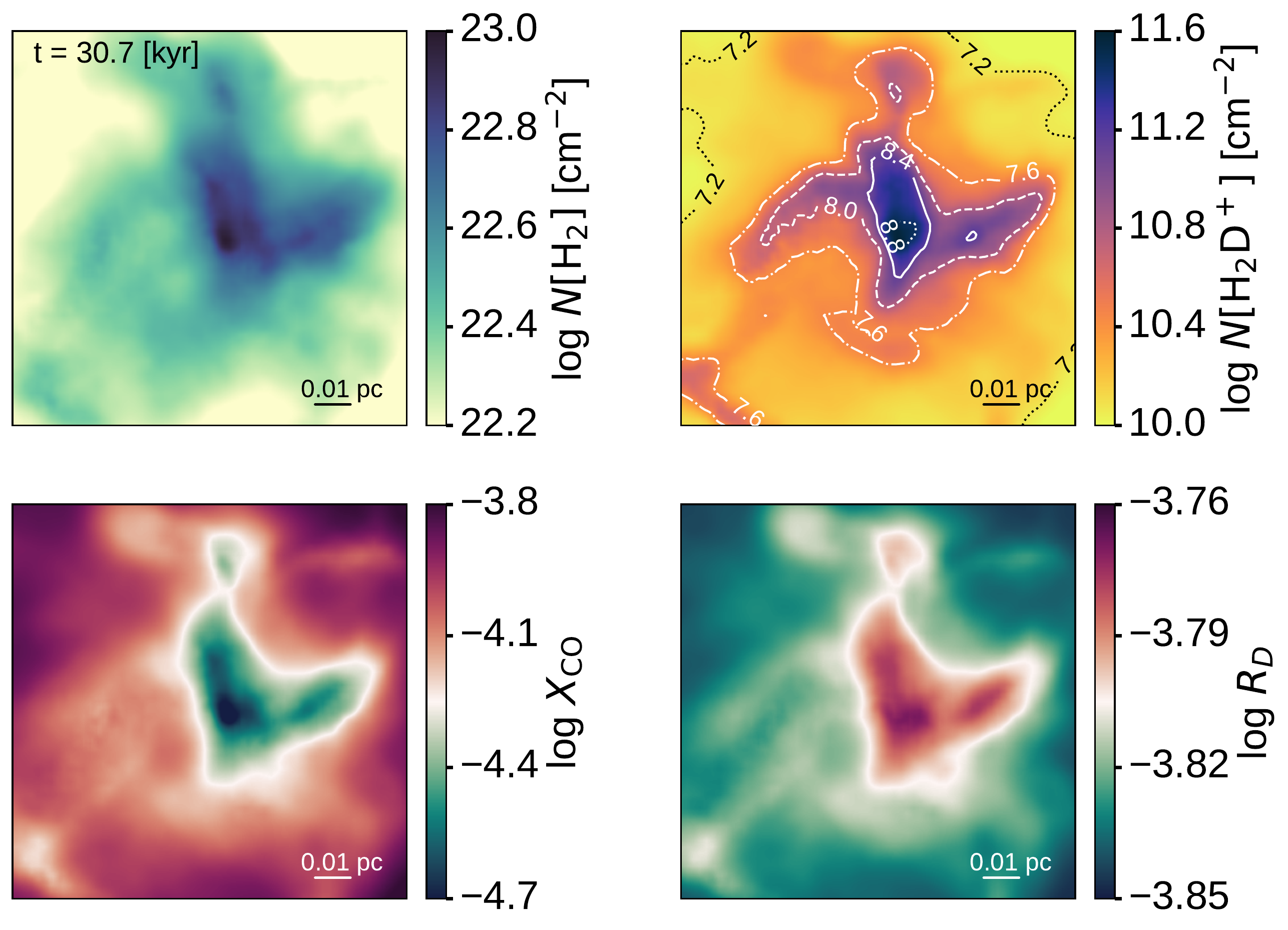}
\caption{Total H$_2$ (top left), $R_D$ (bottom right), and $X_\mathrm{CO}$ (bottom left) maps for our reference run at $\sim$30 kyr of evolution. H$_2$D$^+$ map is shown in the top right panel, with the overplotted white contours representing the $N(\mathrm{D}_2\mathrm{H}^+)$. The column densities are obtained on scales of 0.2 parsecs.}\label{fig:map}
\end{figure*}

\begin{figure}
\includegraphics[scale=0.35]{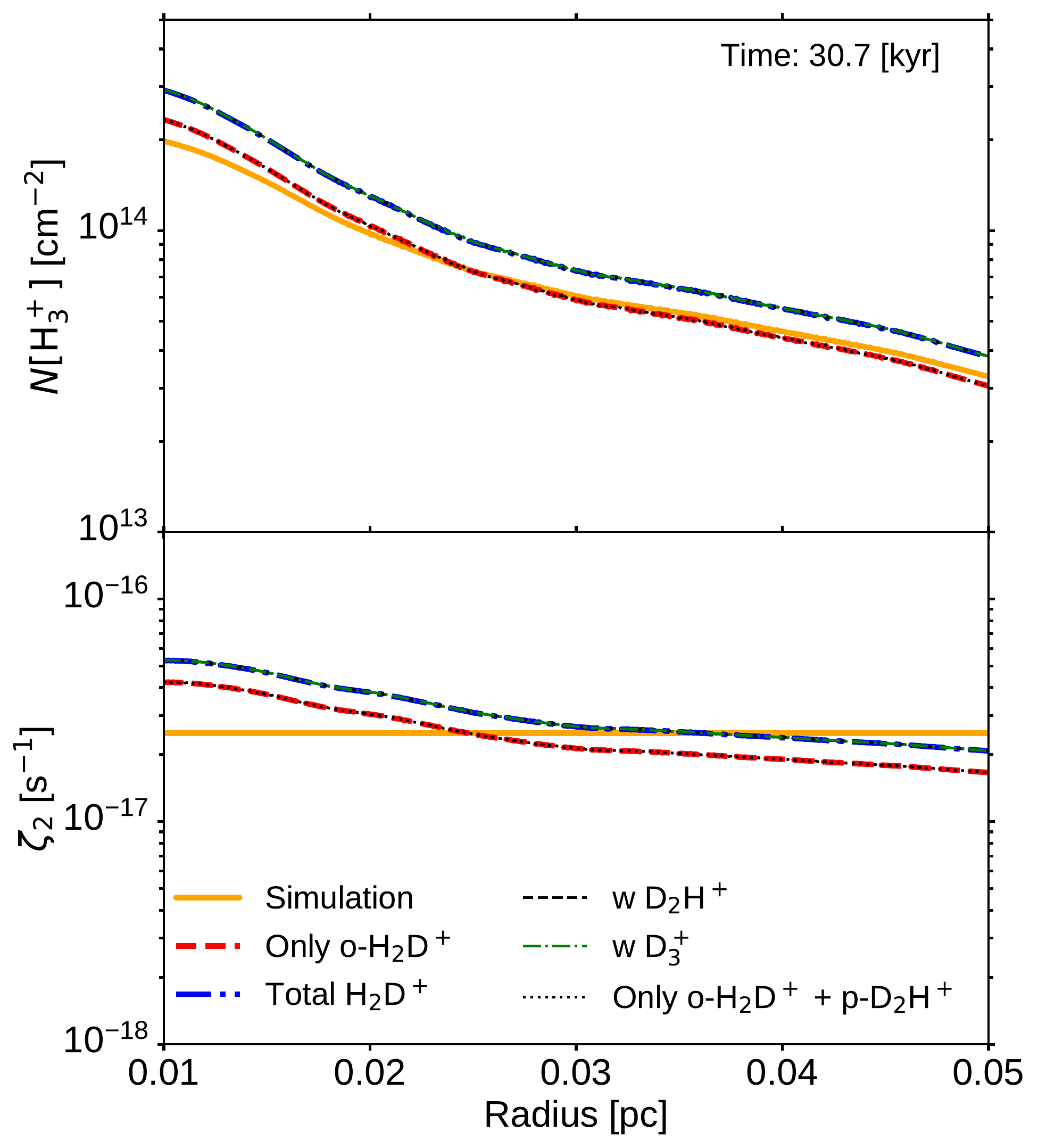}
\caption{Radial profile of the H$_3^+$ column density (top panel) and CRIR (bottom panel) for different analytical approximations, compared to the values in the simulation. The results are shown after 30 kyr of evolution. We show different approximations depending on how many isotopologues have been included in Eq.\ref{eq:h3p}.}\label{fig:profileH3}
\end{figure}

\begin{figure}
\includegraphics[scale=0.35]{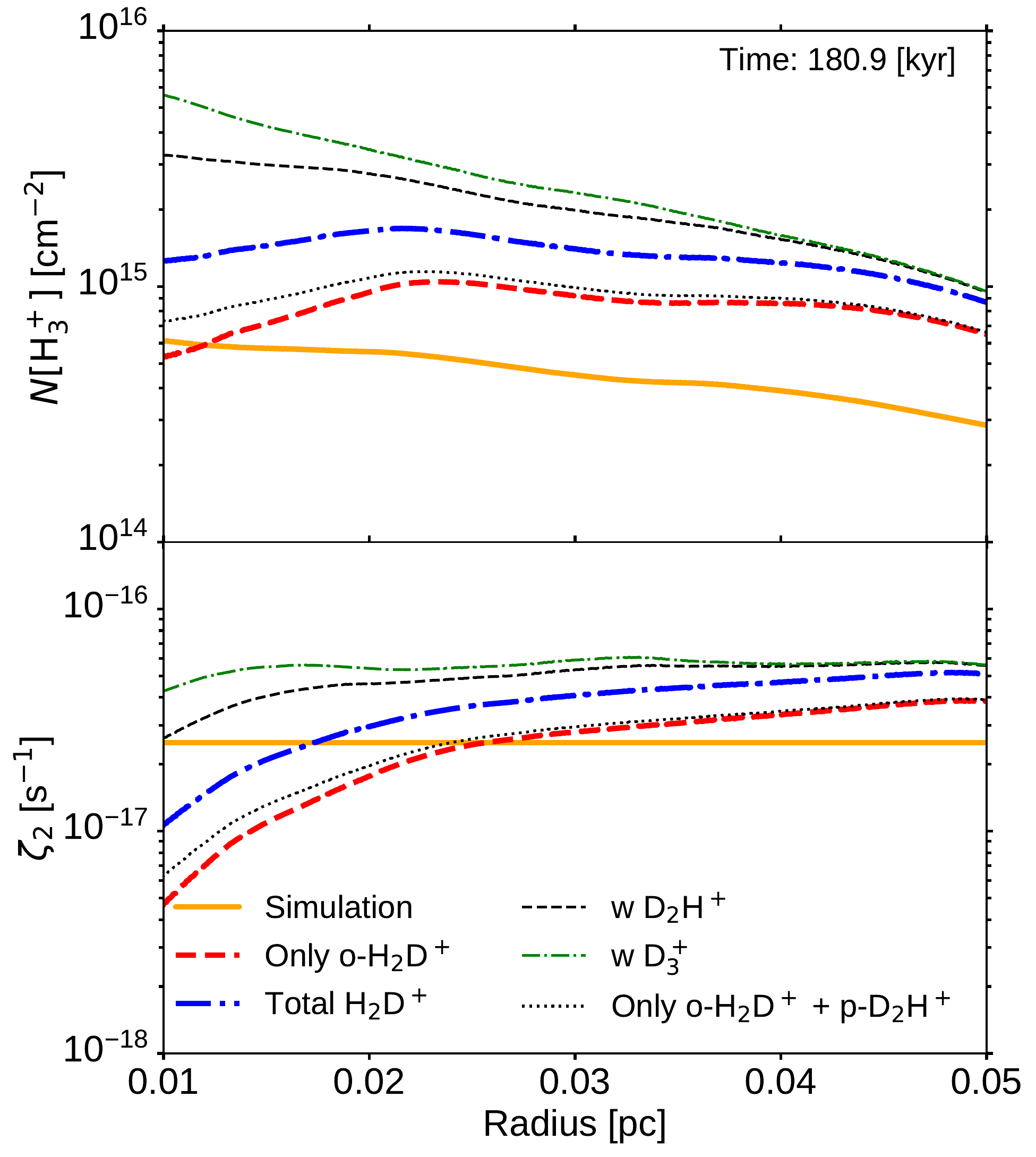}
\caption{Same as Fig.\ref{fig:profileH3} but after 180 kyr.}\label{fig:profile1}
\end{figure}

\section{Results}

%\subsection{Simulations}
%\paragraph{Hydrodynamics}
We use the suite of simulations of collapsing high-mass cores and clumps presented by \citet{Bovino2019}, plus a new set of simulations where we change the CRIR.
The simulations were performed with \textsc{gizmo} \citep{Hopkins15}, assuming an isothermal ($T = 15$ K) Bonnor-Ebert (BE) sphere as initial condition. We account for magnetic fields, turbulence, and detailed chemistry, with C-N-O bearing species including N$_2$H$^+$, CO, HCO$^+$ and their deuterated forms, and adsorption and desorption processes. The mass and spatial resolution of our simulations are 2$\times$10$^{-4}$ M$_\odot$ and  10$^{-4}$ pc, respectively, and all the realizations showed fast deuterium enrichment, in particular for H$_3^+$ and N$_2$H$^+$, and high levels of CO depletion ($f_\mathrm{dep}\sim 20-500$).

%We will employ these simulations to validate analytical formulae to retrieve the CRIR as reported in the next section. To build a robust case, we will do this in time, radially, and for different chemical and dynamical initial conditions. 
In our simulations, the CRIR is set to a constant value, and it is varied to assess how it affects the deuteration timescale and the depletion process. We have selected a reference core of 20 M$_\odot$,  with a size of 0.17 pc, supervirial ($\alpha_\mathrm{vir} = 4.32$), with an average density of $\langle n \rangle = 2.2 \times 10^4$ cm$^{-3}$. This is what we call a "slow collapse" core, highly supported by turbulence and magnetic fields with an average free-fall time of 260 kyr. Maps of the column densities of the main tracers are shown in Fig. \ref{fig:map} after 30 kyr of evolution.  
%We have selected early times because D$_2$H$^+$ and D$_3^+$ are not yet formed in relevant amounts (column densities are at least two orders of magnitude lower compared to H$_2$D$^+$), and our analytical approach provides better results. 
At this stage, the core is slowly collapsing, supported by turbulence, and it is starting to fragment. Column densities of D$_2$H$^+$ and D$_3^+$  are at least two orders of magnitude lower compared to H$_2$D$^+$. The deuteration fraction of HCO$^+$ (i.e. $R_\mathrm{D}$) is on average around 10$^{-3}$, while CO is starting to freeze-out ($f_\mathrm{dep}\sim 5$), in particular in the center of the core. These represent the quantities that we plug into Eq. \ref{eq:h3p}--\ref{eq:crs} to estimate the CRIR.

In Fig.~\ref{fig:profileH3}, we report the H$_3^+$ column density profile obtained from the simulation in comparison with the one obtained from our analytical formula (Eq. \ref{eq:h3p}) on a scale of 0.05 pc for the different approximations.

% i) the full formula, ii) the case where we do not include D$_3^+$ as it is not observable, and iii) a case with the tracers which are commonly observed with ALMA and/or APEX. 
As D$_3^+$ and D$_2$H$^+$ are not yet formed in relevant amounts, the effect of including them in D$[\mathrm{H}_3^+]$ or not is not dramatic. The main difference between the approximations comes from the inclusion (or not) of the \textit{para} form of H$_2$D$^+$ which at this stage is the only relevant isotopologue together with its \textit{ortho} counterpart. Compared to the column density obtained from simulations, we notice a very small error whatever the assumption we employ. This is also reflected in the CRIR profile (bottom panel of Fig.~\ref{fig:profileH3}) where we are able to recover the CRIR set in the simulation with our analytical formula within a factor of less than two depending on the istopologues that we include.

While the analytical formula is working very well at early dynamical times, the error increases as we proceed in time. This is highlighted in Fig.~\ref{fig:profile1}, where we report the H$_3^+$ column density and the CRIR at $\sim$180 kyr. We see that the inferred column density is now affected by a larger error, on average it is overstimated by a factor of five. The overestimate is not surprising considering that we are neglecting important destruction paths for H$_3^+$ (e.g. O and e$^-$). However, when we look at the estimate of the CRIR the final error is lower, and still depends on the formula employed and on the spatial scale. 

%Considering that we are using column densities, which have been shown to carry only a very small fraction of the real amount \citep[see][]{Bovino2019}, it is clear that we are diluting or smearing out discrepancies.

\paragraph*{Error estimates} Because of the significant variations in the CRIR estimate, both in space and time, the best approach to get an error estimate is via a statistical analysis of the results, averaged over time and over different spatial scales. We proceed in two steps: i) we first construct a temporal evolution of the mean CRIR estimated from the average column densities over different spatial scales $\theta$ ranging in between 0.02-0.05~pc, and then ii) average each CRIR history in time to obtain a mean value and its standard deviation. Finally, we estimate a single representative CRIR for each approximation by applying a \textit{weighted mean} to the different $\theta$, defined as 

\begin{equation}
	\bar{\zeta}_2 = \frac{\sum_{i=1}^{N} w(\theta_i) \langle \zeta_2(\theta_i) \rangle }{\sum_{i=1}^{N} w(\theta_i)}\,,
\label{eq:err1}
\end{equation}
\noindent where $\langle \zeta_2(\theta_i) \rangle$ is the time average on the different spatial scales. Assuming that each spatial scale gives a statistically independent measure of the CRIR, the weights can be defined as $w(\theta_i)~=~1 / \sigma^2(\theta_i)$, being $\sigma(\theta_i)$ the standard deviation for the \textit{i}th scale. The standard error for the weighted mean is then calculated as 

\begin{equation}
	\sigma_{\bar{\zeta}_2} = \left(\sqrt{\sum_{i=1}^{N} w(\theta_i)}\right)^{-1}\,.
\label{eq:err2}
\end{equation}
This is applied to a suite of simulations with different dynamical initial conditions and initial CRIR. Final results are reported in Table \ref{tab:results}, where we specify the type of simulation and the employed CRIR. 
Our estimate of the average CRIR is very close to the real value, with changes within a factor of $\alpha=\zeta_2^\mathrm{real} / \bar{\zeta_2} \sim 0.3-0.8$, with the only exception being the case with a very high CRIR, where the error is still within a factor of two but we notice an inverted trend for some of the formulae (e.g. a factor of 2.7 for the formula where we employ only o-H$_2$D$^+$). 
%The inverse of this factor can  be used as a correction factor in our Eq. \ref{eq:crs} based on the formula we use to evaluate N[H$_3^+$]. 
We calculate the standard deviation of the individual $\alpha$ by propagating the errors and then computing a weighted mean to obtain an average correction factor $\bar{\alpha}$ to be used in Eq.\ref{eq:crs}, with its corresponding uncertainty $\sigma_{\bar{\alpha}}$. The final results are reported in Table \ref{tab:results1} from where we see that the correction factor ranges between 0.4-0.7, i.e. the error is always within a factor of two\footnote{We have also applied the entire procedure by using the median to avoid the effect of outliers, but the final $\bar{\alpha}$ are not affected.}.

%have a final averaged CRIR estimate $\bar{\zeta_2} = \alpha \zeta_2$ with $\alpha$ values of: 1.16 for (1), 0.64 for (2), 0.53 for (3), 0.49 for (4), and 0.88 for (5). As clear from the correction factor, we are very close to the real value, with a maximum error of a factor of two.

%\begin{table}
%	\begin{tabular}{lccc}
%		\hline
%		Formula & $\bar{\zeta_2}$ & $\sigma$ & $\bar{\zeta_2}/\zeta_2^\mathrm{sim}$\\
%		\hline
%		full &  7.4 $\times 10^{-17}$ & 1.2 $\times 10^{-17}$ & 3\\
%		w/o D$_3^+$ & 6.6 $\times 10^{-17}$ & 1.3 $\times 10^{-17}$ & 2.6\\
%		only o-H$_2$D$^+$ \& p-D$_2$H$^+$ & 4.0 $\times 10^{-17}$ & 1.1 $\times 10^{-17}$  & 1.6\\
%		$\zeta_2^\mathrm{sim}$ & 2.5 $\times 10^{-17}$	& -  \\ 
%	\end{tabular}
%	\caption{Radial and time averaged cosmic ray ionisation rate $\bar{\zeta_2}$ obtained from Eq.\ref{eq:crs} with different levels of approximation: \textit{full}, the entire formula including all H$_3^+$ isotopologues, \textit{w/o D$_3^+$}, i.e. including only real observables, in this case all the isomers of H$_2$D$^+$ and D$_2$H$^+$, and the last case where we only consider in Eq.\ref{eq:crs} the tracers observable with ALMA or APEX, i.e. o-H$_2$D$^+$ and p-D$_2$H$^+$. We also report the final standard deviation $\sigma$. As a reference the "real" value employed in the simulations is also shown.}\label{tab:results}
%\end{table} 
\begin{table}
    \centering
    \caption{The \textit{Method} column refers to which species are used to calculate D$[\rm H_3^+]$ in Eq.\ref{eq:h3p}. (1): only o-H$_2$D$^+$, (2):  total $\rm{H}_2D^+$, (3): the observable isotopologues ($\rm{H}_2\rm{D}^+$ and $\mathrm{D}_2\mathrm{H}^+$), (4):  all the isotopologues, i.e., we also include D$_3^+$, and (5): only the species observable with ALMA/APEX (o-$\mathrm{H}_2\mathrm{D}^+$ and p-D$_2$H$^+$). The last two columns report the correction factor $\alpha$ and its uncertainty. Details on the different simulations can be found in \citet{Bovino2019}. The H$_2$ ortho-to-para ratio (OPR) is set to 3 unless specified.}
    \label{tab:results}
    \begin{tabular}{ccccc}
	%\hline
	\hline	
	    Method & $\bar{\zeta_2}$ & $\sigma_{\bar{\zeta_2}}$ &  $\alpha$ & $\sigma_{\alpha}$\\
	    \hline
        \multicolumn{5}{l}{\textbf{M0 Fast collapse core}: $\zeta_{\rm 2} = 2.5\times 10^{-17}$ s$^{-1}$:}\\
        (1) & $3.3466\times10^{-17}$ & $5.7996\times10^{-18}$ & $0.7470$ & $0.1295$ \\
        (2) & $6.2106\times10^{-17}$ & $1.1092\times10^{-17}$ & $0.4025$ & $0.0719$ \\
        (3) & $6.2418\times10^{-17}$ & $1.1162\times10^{-17}$ & $0.4005$ & $0.0716$ \\
        (4) & $6.2418\times10^{-17}$ & $1.1162\times10^{-17}$ & $0.4005$ & $0.0716$ \\
        (5) & $4.9622\times10^{-17}$ & $8.8558\times10^{-18}$ & $0.5038$ & $0.0899$ \\
        \multicolumn{3}{l}{}\\
        \multicolumn{5}{l}{\textbf{M1 Slow collapse core}, $\zeta_{\rm 2} = 2.5\times 10^{-17}$ s$^{-1}$:}\\
        (1) & $3.1568\times10^{-17}$ & $9.8170\times10^{-18}$ & $0.7919$ & $0.2463$ \\
        (2) & $5.3419\times10^{-17}$ & $1.4347\times10^{-17}$ & $0.4680$ & $0.1257$ \\
        (3) & $6.5516\times10^{-17}$ & $1.2758\times10^{-17}$ & $0.3816$ & $0.0743$ \\
        (4) & $7.3836\times10^{-17}$ & $1.2076\times10^{-17}$ & $0.3386$ & $0.0554$ \\
        (5) & $3.9854\times10^{-17}$ & $1.1719\times10^{-17}$ & $0.6273$ & $0.1844$ \\
        \multicolumn{3}{l}{}\\
        \multicolumn{5}{l}{\textbf{M1 Slow collapse core}, $\zeta_{\rm 2} = 2.5\times 10^{-18}$ s$^{-1}$:}\\
        (1) & $3.6303\times10^{-18}$ & $6.1257\times10^{-19}$ & $0.6886$ & $0.1162$ \\
        (2) & $5.6742\times10^{-18}$ & $9.4603\times10^{-19}$ & $0.4406$ & $0.0735$ \\
        (3) & $5.6976\times10^{-18}$ & $9.4811\times10^{-19}$ & $0.4388$ & $0.0730$ \\
        (4) & $5.6977\times10^{-18}$ & $9.4811\times10^{-19}$ & $0.4388$ & $0.0730$ \\
        (5) & $4.5361\times10^{-18}$ & $7.5718\times10^{-19}$ & $0.5511$ & $0.0920$ \\
        \multicolumn{3}{l}{}\\
        \multicolumn{5}{l}{\textbf{M1 Slow collapse core}, $\zeta_{\rm 2} = 2.5\times 10^{-16}$ s$^{-1}$:}\\
        (1) & $9.2224\times10^{-17}$ & $6.1415\times10^{-17}$ & $2.7110$ & $1.8050$ \\
        (2) & $1.7484\times10^{-16}$ & $1.0019\times10^{-16}$ & $1.4300$ & $0.8194$ \\
        (3) & $2.6436\times10^{-16}$ & $1.3270\times10^{-16}$ & $0.9457$ & $0.4747$ \\
        (4) & $3.0737\times10^{-16}$ & $1.4427\times10^{-16}$ & $0.8134$ & $0.3818$ \\
        (5) & $1.1621\times10^{-16}$ & $7.2152\times10^{-17}$ & $2.1510$ & $1.3360$ \\
        \multicolumn{3}{l}{}\\
        \multicolumn{5}{l}{\textbf{M1 Slow collapse core}, $\zeta_{\rm 2} = 2.5\times 10^{-17}$ s$^{-1}$,\, OPR(H$_2$) = 0.1:}\\
        (1) & $2.3604\times10^{-17}$ & $8.7352\times10^{-18}$ & $1.0590$ & $0.3920$ \\
        (2) & $4.3713\times10^{-17}$ & $1.3791\times10^{-17}$ & $0.5719$ & $0.1804$ \\
        (3) & $6.2305\times10^{-17}$ & $1.5183\times10^{-17}$ & $0.4013$ & $0.0978$ \\
        (4) & $7.3752\times10^{-17}$ & $1.4537\times10^{-17}$ & $0.3390$ & $0.0668$ \\
        (5) & $3.0594\times10^{-17}$ & $1.0435\times10^{-17}$ & $0.8172$ & $0.2787$ \\
        \multicolumn{3}{l}{}\\
        \multicolumn{5}{l}{\textbf{M3 Fragmented clump: $\zeta_{\rm 2} = 2.5\times 10^{-17}$ s$^{-1}$:}}\\
        (1) & $2.7583\times10^{-17}$ & $5.1299\times10^{-18}$ & $0.9064$ & $0.1686$ \\
        (2) & $4.9326\times10^{-17}$ & $9.2854\times10^{-18}$ & $0.5068$ & $0.0954$ \\
        (3) & $5.0298\times10^{-17}$ & $9.5295\times10^{-18}$ & $0.4970$ & $0.0942$ \\
        (4) & $5.0336\times10^{-17}$ & $9.5365\times10^{-18}$ & $0.4967$ & $0.0941$ \\
        (5) & $3.9064\times10^{-17}$ & $7.3345\times10^{-18}$ & $0.6400$ & $0.1202$ \\
        \hline
    \end{tabular}
\end{table}

\begin{figure}
\includegraphics[scale=0.5]{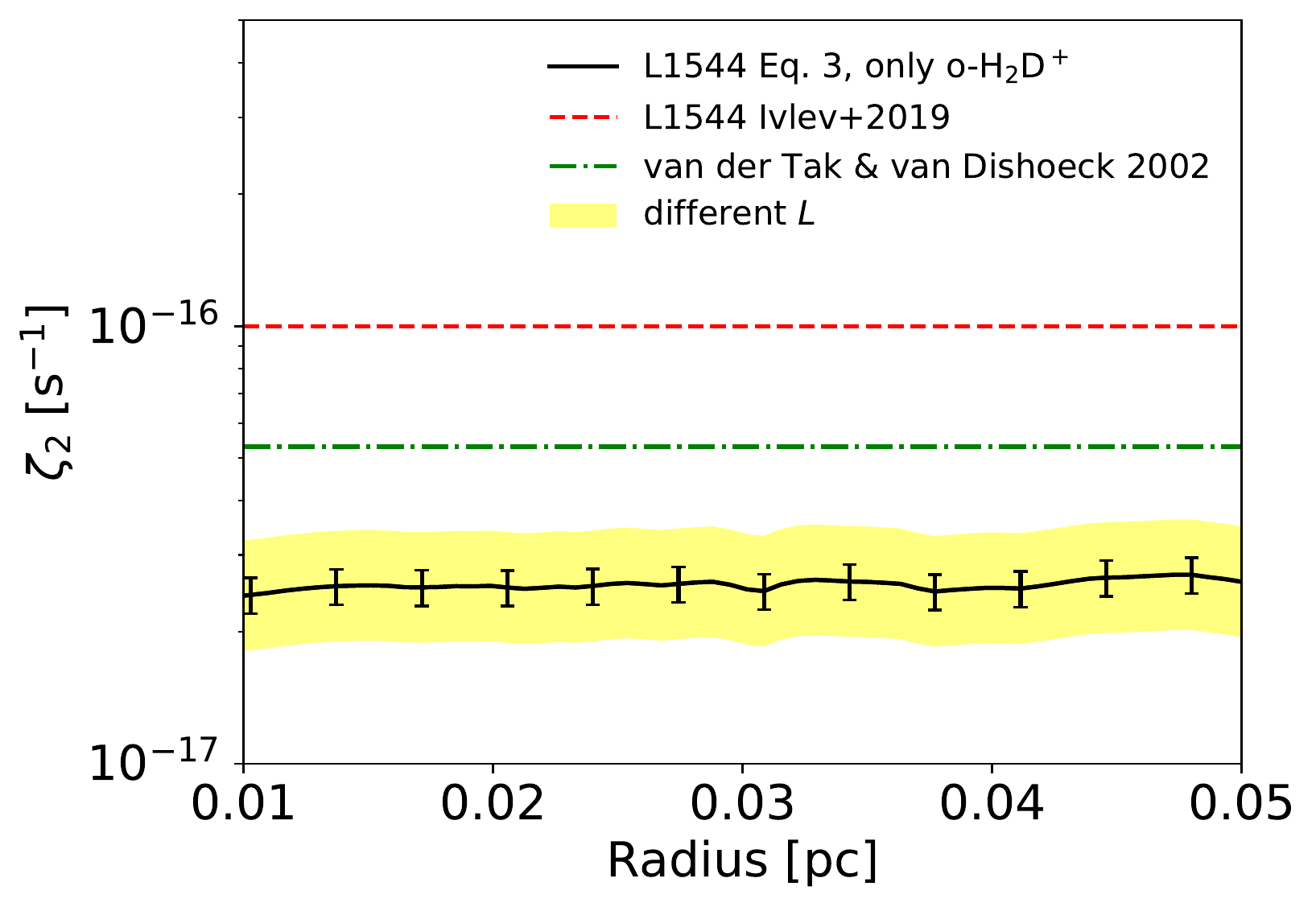}
\caption{Cosmic-ray ionisation rate from Eq. \ref{eq:crs}, by employing the available column density profiles from \citet{Vastel2006}. The green dotted-dashed line is $\zeta_\mathrm{2} = 5.3 \times 10^{-17}$ s$^{-1}$ reported by \citet{vanderTak2000}. The dashed red line instead represent recent estimates reported by \citet{Ivlev2019} for the same source. We have assumed a path length of 0.034 pc which is representative of the extension of o-H$_2$D$^+$ emission. The shaded area represents the variability of our results with $L$.}\label{fig:l1544}
\end{figure}

\paragraph*{Application to L1544} To test our analytical approach, we have selected one of the best studied low-mass cores, L1544, and used data from \citet{Vastel2006} who provide in their Fig.~6 the radial profiles of the column densities for different tracers.
%as a function of distance from the center of the core. 
With this data in hand, we computed the CRIR as shown in Fig.~\ref{fig:l1544}. We estimate the path length by taking the geometrical average between the major and the minor axis of the \mbox{o-H$_2$D$^+$} emission obtained by fitting the 50\% contour with an ellipse (see their Fig. 4). We obtain $L = 0.034$~pc and $\zeta_2 \sim$ 2.5$\times 10^{-17}$~s$^{-1}$. However, we have to consider that the three-dimensional shape of the core is not known, therefore we also report the cases for $L$ in between the major and the minor axis (yellow shaded area in Fig.\ref{fig:l1544}). The CRIR moves to lower (higher) values with peaks around 3$\times 10^{-17}$ s$^{-1}$. Overall, our estimated CRIR obtained from an unconstrained and a model-independent approach, is close to typical values reported by \citet{vanderTak2000}\footnote{We are using a CRIR per hydrogen molecule instead of per hydrogen atom as reported in the former paper, then there is a factor of 2.3 difference.} but far from the value of $\zeta_2 \sim 10^{-16}$ s$^{-1}$ measured in diffuse clouds and recently reported by \citet{Ivlev2019} for the same pre-stellar core. Thus, chemistry appears to favor lower values of $\zeta_2$ in dense cores\footnote{To reach $\zeta_2 \sim 10^{-16}$ s$^{-1}$ $L$ should be significantly smaller than the observed o-H$_2$D$^+$ emission toward L1544.}.

 \begin{table}
    \centering
    \caption{Error propagation from averaging over the different simulations. The different column refer to the different equations we are using for D$[\rm H_3^+]$.}\label{tab:results1}
    \begin{tabular}{lccccc}
    \hline
    %\hline
    & (1) & (2) & (3) & (4) & (5)\\
    \hline
	$\bar{\alpha}$ & $0.7712$ & $0.4528$ & $0.4223$ &  $0.3881$ &$0.5705$\\
	$\sigma_{\bar{\alpha}}$ & $0.0721$ & $0.0414$ & $0.0357$ & $0.0308$& $0.0532$\\
	\hline
  \end{tabular}
\end{table}

\section{Discussion and Conclusions}
In this Letter we propose and test an analytical approach to estimate the CRIR in the densest regions of molecular clouds. We have built the formula from first principles, under steady-state conditions, by following pioneering works in this field. The new idea behind the presented approach is based on two main components: i) the estimate of the H$_3^+$ column density from its deuterated isotopologues, and ii) the validation of the methodology via three-dimensional high-resolution simulations. This allowed us to explore different approximations and to test the statistical relevance of the approach by running an error analysis on different realizations. Overall, our method is providing small errors and a deviation from the real value of maximum a factor of two. 
It is worth noting that the applicability of the formula depends on the availability of high-resolution observations being constrained by spatial scales (validity for $R \leq 0.05$ pc), in particular for CO that can be affected by freeze-out, and the H$_3^+$ isotopologues. In this respect, we propose different versions of the analytical formula to give more flexibility on its usage. 
%These depend on a combination of different isotopologues for the estimate of the H$_3^+$ column density. In particular, the mix of o-H$_2$D$^+$ and p-D$_2$H$^+$ allows to apply the methodology by simply using the two tracers which are easier to observe with APEX or ALMA.

%We stress that the need for high-resolution observations is crucial since the applicability of our methodology is constrained by spatial scales, i.e. it could be applied only to subparsec scales.

To further validate our results, we have applied the analytical formula to the well-studied low-mass core L1544, for which observations of the needed tracers exist, and we have found an average value for the CRIR of 2 -- 3 $\times 10^{-17}$ s$^{-1}$. 
%This compares very well with old estimates obtained by \citet{vanderTak2000} ($\zeta_2 = 5 \times 10^{-17}$ s$^{-1}$).
%and new ones based on dust properties reported by \citet{Ivlev2019}, $\zeta_2 \sim 10^{-16}$ s$^{-1}$.

A very critical point of our approach is  related to the path length $L$. While theoretically we can provide a precise number for $L$, observationally this is an arbitrary choice. In \citet{McCall1999}\, $L$ is obtained from a mix of observations and models as the ratio between the column and the number density of H$_3^+$. The column length can also be provided by the extension of the emission of the involved tracers, or by estimates of the size of the core. 
Considering the error analysis that we have performed on results which we can consider very robust, we can state that the main source of error in the presented approach comes from the choice of the path length $L$.

To conclude, even considering the uncertainties coming from the choice of the path length, and the intrinsic error which affects our approach, in particular the steady-state assumption, Eq. \ref{eq:crs} represents the first attempt of providing a robust tool to evaluate the cosmic-ray ionisation rate in dense cores, a parameter that has a strong impact on theoretical models, and on our understanding of star-forming regions.

\section*{Acknowledgements}
%We are grateful to Sandro Villanova for useful discussion on the error analysis. 
SB is financially supported by CONICYT Fondecyt Iniciaci\'on (project 11170268). SB and DRGS are supported by CONICYT programa de Astronomia Fondo Quimal 2017 QUIMAL170001. AL acknowledges support by the European Research Council No. 740120 `INTERSTELLAR'.
%DRGS thanks for funds through CONICYT PIA ACT172033 and CONICYT Fondecyt regular (project 1161247).  
SFC is supported by CONICYT Programa de Astronom\'ia Fondo ALMA-CONICYT 2017 Project \#31170002.
%The simulations were performed with resources provided by the \textit{KULTRUN Astronomy Hybrid Cluster} at Universidad de Concepci\'on.

%%%%%%%%%%%%%%%%%%%%%%%%%%%%%%%%%%%%%%%%%%%%%%%%%%

%%%%%%%%%%%%%%%%%%%% REFERENCES %%%%%%%%%%%%%%%%%%

% The best way to enter references is to use BibTeX:

\bibliographystyle{mnras}
\bibliography{mybib} % if your bibtex file is called example.bib

%%%%%%%%%%%%%%%%%%%%%%%%%%%%%%%%%%%%%%%%%%%%%%%%%%

%%%%%%%%%%%%%%%%% APPENDICES %%%%%%%%%%%%%%%%%%%%%

%%%%%%%%%%%%%%%%%%%%%%%%%%%%%%%%%%%%%%%%%%%%%%%%%%

% Don't change these lines
\bsp	% typesetting comment
\label{lastpage}
\end{document}